\documentstyle[12pt]{article}

\input epsf.sty
\newcommand{\inclfig}[2]{}
\newcommand{\insertfig}[2]{}

\newcommand{\Bx}{x_{\rm B}}
\newcommand{\ft}[2]{{\textstyle\frac{#1}{#2}}}

\begin{document}

\begin{titlepage}

\centerline{\large\bf Twist-three observables in}
\centerline{\large\bf deeply virtual Compton scattering on the nucleon.}

\vspace{15mm}

\centerline{\bf A.V. Belitsky$^{a,c}$, A. Kirchner$^b$, D. M\"uller$^{b,c}$,
                A. Sch\"afer$^b$}

\vspace{10mm}

\centerline{\it $^a$C.N.\ Yang Institute for Theoretical Physics}
\centerline{\it State University of New York at Stony Brook}
\centerline{\it NY 11794-3840, Stony Brook, USA}

\vspace{5mm}

\centerline{\it $^b$Institut f\"ur Theoretische Physik,
                Universit\"at Regensburg}
\centerline{\it D-93040 Regensburg, Germany}

\vspace{5mm}

\centerline{\it $^c$Fachbereich Physik,
                Universit\"at  Wuppertal}
\centerline{\it D-42097 Wuppertal, Germany}

\vspace{15mm}

\centerline{\bf Abstract}

\vspace{0.5cm}

\noindent

We study twist-three effects in deeply virtual Compton scattering on an
unpolarized spin-1/2 target. A careful definition of observables as Fourier
moments w.r.t.\ the azimuthal angle allows for a clear separation of
twist-two and -three effects. Although the latter are power suppressed, they
give leading contributions to the twist-three asymmetries and do not affect
the twist-two observables.

\vspace{5cm}


\vspace{0.5cm}


\end{titlepage}


{\bf 1.} The understanding of the hadron substructure as well as the
dynamics of hadron constituents requires mutual efforts from both the
theoretical and experimental side. Theoretically, we have to deduce as much
information as possible from perturbative and non-perturbative QCD
calculations. Until now, on the experimental side the nucleon structure has
been mostly probed in inclusive processes, e.g.\ measurements of
deep-inelastic structure functions in leptoproduction experiments, which
give access to parton distribution functions. Additional information has
been gained from exclusive measurements of electroweak form factors
sensitive at high momentum transfer to the lowest Fock component of the
hadron wave function. Since recently it became clear that far richer
information can be extracted from generalized parton distributions (GPDs)
\cite{MulRobGeyDitHor94,Ji96,Rad96}, which appear e.g.\ in the processes of
deeply virtual Compton scattering (DVCS) $\ell N \to \ell'\gamma N'$ and exclusive
leptoproduction of mesons $\ell N \to \ell' M N'$. The GPDs include as limits
the conventional hadron characteristics alluded to before
\cite{MulRobGeyDitHor94,Ji96,Rad96}.

The first experimental results \cite{Ama00,Fav00,Jlab} show the
experimental feasibility of DVCS measurements. The DVCS process
contributes together with the contaminating Bethe-Heitler (BH) one to
exclusive electroproduction of the real photon and requires detailed
theoretical studies. Obviously, the first issue is to define
appropriate observables and to determine the phase space regions that would
allow for a clean interpretation of experimental measurements. As
has been shown in previous studies, diverse asymmetries give an
access to the interference term. Here the DVCS signal gets augmented by the
Bethe-Heitler process, which provides the required handle on the GPDs from
leptoproduction reactions \cite{Ji96,GouDiePirRal97,BelMueNieSch00}.
Certainly, at very large values of the photon virtuality ${\cal Q}^2$ the
DVCS process is dominated by its leading twist-two approximation arising
from the so-called handbag diagram (including radiative corrections).
Practically the onset of scaling can only be judged experimentally as
present day theory does not provide necessary non-perturbative tools
for the computation of soft contributions to such a reaction. For deep-inelastic
scattering (DIS) it turned out, that the twist-two approximation is valid down
to a rather low photon virtuality of order of a few GeV. Although the DIS process
is given by the absorptive part of forward Compton scattering, calculated by
means of the same operator product expansion, the DVCS case remains an issue
for studies. Since the average ${\cal Q}^2$ for all present facilities is
about 4 GeV$^2$ or even below, it is vital for a clean isolation of the leading
twist GPDs, to have a control over power suppressed effects.

Recently, the issues of twist-three effects
\cite{PenPolShuStr00}--
\cite{BalLaz01} and target mass corrections
\cite{BelMue01} have been addressed in the literature (in the latter case
only partially). In the present paper we make a step towards accomplishing
the goal of handling the higher twist contributions by defining and
computing twist-three observables in the DVCS cross section with polarized
lepton beam and unpolarized nucleon target.

{\bf 2.} The DVCS hadronic tensor is given by the time ordered product of the
electromagnetic currents $j_\mu = \bar\psi \gamma_\mu \psi$ sandwiched
between hadronic states with different momenta. In leading order of
perturbation theory it reads \cite{BelMul00b}
\begin{eqnarray}
\label{HadronicTensor}
T_{\mu\nu} (q, P, \Delta)
\!\!\!&=&\!\!\!
i \int dx {\rm e}^{i x \cdot q}
\langle P_2 | T j_\mu (x/2) j_\nu (-x/2) | P_1 \rangle \\
&=&\!\!\!
- {\cal P}_{\mu\sigma} g_{\sigma\tau} {\cal P}_{\tau\nu}
\frac{q \cdot V_1}{P \cdot q}
+ \left( {\cal P}_{\mu\sigma} P_\sigma  {\cal P}_{\rho\nu}
+ {\cal P}_{\mu\rho}  P_\sigma {\cal P}_{\sigma\nu} \right)
\frac{V_{2\, \rho}}{P \cdot q}
- {\cal P}_{\mu\sigma} i\epsilon_{\sigma \tau q \rho} {\cal P}_{\tau\nu}
\frac{A_{1\, \rho}}{P \cdot q} , \nonumber
\end{eqnarray}
where we have kept all contributions to twist-three accuracy. Here we have
used conventional conditions on invariants in the Bjorken limit, $- q^2 \sim
P \cdot q =$ large, $\Delta^2 =$ small, $\xi = -q^2/ q \cdot P =$ fixed,
constructed from the vectors $P = P_1 + P_2$, $\Delta = P_2 - P_1$, and $q =
(q_1 + q_2)/2$. The vectors $P_1$ ($q_1$) and $P_2$ ($q_2$) refer here to
the incoming and outgoing proton (photon) momentum, respectively. The reality
of the final state photon implies the presence of only one scaling
variable $\xi$. Current conservation in the tensor decomposition
(\ref{HadronicTensor}) is ensured here by means of the projection operator
\begin{equation}
{\cal P}_{\mu\nu} = g_{\mu\nu} - \frac{q_{1 \mu} q_{2 \nu}}{q_1 \cdot q_2} .
\end{equation}
This is consistent with an explicit calculation of the amplitude
(\ref{HadronicTensor}) to twist-three accuracy
\cite{PenPolShuStr00,BelMul00b}. In the above equation the vector $V_{2 \,
\rho}$ is expressed in terms of the vector $V_{1 \, \rho}$ and the
axial-vector $A_{1 \, \rho}$ form factors
\begin{eqnarray}
\label{V2}
V_{2 \, \rho} = \xi V_{1 \, \rho} - \frac{\xi}{2}
\frac{P_\rho}{P\cdot q} q\cdot V_{1} + \frac{i}{2}
\frac{\epsilon_{\rho\sigma\Delta q}}{P\cdot q} A_{1 \, \sigma} .
\end{eqnarray}
The latter are given as convolutions w.r.t.\ the momentum fraction $x$,
$\otimes \equiv \int_{-1}^{1} d x$,
\begin{equation}
\label{Convolution}
V_{1 \, \rho} = C^{(-)} (\xi, x) \otimes v_{1 \, \rho} (x, \xi, \Delta^2) ,
\qquad
A_{1 \, \rho} = C^{(-)} (\xi, x) \otimes a_{1 \, \rho} (x, \xi, \Delta^2)
\end{equation}
of the leading order coefficient functions
\begin{equation}
\xi\, {C_i^{(\mp)}} \left( \xi, x \right)
= Q^2_i (1 - x/\xi - i \epsilon)^{-1} \mp (x \to -x) ,
\end{equation}
with $-$ ($+$) standing for the parity even (odd) case, $Q_i$ being the
electric charge of a quark of type-$i$, and the Fourier transforms of light-ray
operators ($n^2 = 0$) being defined by
\begin{equation}
\left\{
\begin{array}{c}
v_{1 \, \rho}
\\
a_{1 \, \rho}
\end{array}
\right\}(x, \xi, \Delta^2)
=
\int \frac{d \kappa}{2 \pi}
{\rm e}^{i x \kappa (n\cdot P)}
\langle P_2 |
\bar \psi (- \kappa n)
\left\{
\begin{array}{c}
\gamma_\rho
\\
\gamma_\rho \gamma_5
\end{array}
\right\}
[- \kappa, \kappa]
\psi (\kappa n)
| P_1 \rangle .
\end{equation}
The general decomposition of the vector and axial-vector amplitudes, in a
complete basis of form factors to twist-three accuracy, reads
\begin{eqnarray}
\label{Def-V1}
V_{1\, \rho}\!\!\!&=&\!\!\!
P_\rho  \frac{q\cdot h}{q\cdot P} {\cal H}
+
P_\rho  \frac{q\cdot e}{q\cdot P} {\cal E}
+
\Delta_\rho^\perp  \frac{q\cdot h}{q\cdot P}  {\cal H}^3_+
+
\Delta_\rho^\perp  \frac{q\cdot e}{q\cdot P}  {\cal E}^3_+
+
\widetilde{\Delta}_\rho^\perp
\frac{q\cdot \widetilde{h}}{q\cdot P} \widetilde{{\cal H}}^3_-
+
\widetilde{\Delta}_\rho^\perp
\frac{q\cdot \widetilde{e}}{q\cdot P} \widetilde{{\cal E}}^3_-
,\\
\label{Def-A1}
A_{1\, \rho}\!\!\!&=&\!\!\!
P_\rho  \frac{q\cdot \widetilde h}{q\cdot P} \widetilde{\cal H}
+
P_\rho  \frac{q\cdot \widetilde e}{q\cdot P} \widetilde{\cal E}
+
\Delta_\rho^\perp  \frac{q\cdot \widetilde h}{q\cdot P}
{\cal \widetilde H}^3_+
+ \Delta_\rho^\perp  \frac{q\cdot \widetilde e}{q\cdot P}
{\cal \widetilde E}^3_+
+
\widetilde{\Delta}_\rho^\perp  \frac{q\cdot h}{q\cdot P}
{\cal H}^3_-
+
\widetilde{\Delta}_\rho^\perp  \frac{q\cdot e}{q\cdot P}
{\cal E}^3_- ,
\end{eqnarray}
where $\Delta_\rho^\perp \equiv \Delta_\rho + \xi P_\rho$,
$\widetilde\Delta_\rho^\perp \equiv i \epsilon_{\rho\Delta P q}/{P\cdot q}$
and the Dirac bilinears are defined conventionally by
\begin{eqnarray*}
&& h_{\rho}
= \bar U (P_2, S_2) \gamma_\rho U (P_1, S_1),
\hspace{1.2cm}
e_{\rho} =
\bar U (P_2, S_2) i \sigma_{\rho\sigma} \frac{\Delta_\sigma}{2 M} U (P_1, S_1),
\nonumber\\
&&\widetilde{h}_\rho
= \bar U (P_2, S_2) \gamma_\rho \gamma_5  U (P_1, S_1),
\qquad
\widetilde e_{\rho}
=  \frac{\Delta_\rho}{2 M} \bar U (P_2, S_2) \gamma_5  U(P_1, S_1) ,
\end{eqnarray*}
with $U$ being the nucleon bispinor.

In order to simplify our presentation, we introduce a unified convention
for the twist-two ${\cal F} \equiv \{ {\cal H}, {\cal E}, \widetilde {\cal
H}, \widetilde {\cal E}\}$ and twist-three ${\cal F}^3_\pm \equiv \{ {\cal
H}^3_\pm, \dots, \widetilde {\cal E}^3_\pm\}$ Compton form factors.
Replacing the sets of Compton form factors ${\cal F}$ and ${\cal F}^3_\pm$
by the sets of GPDs $F_i\equiv \{ H_i, \dots, \widetilde E_i \}$ and so on,
we write in analogy to Eqs.\ (\ref{Def-V1},\ref{Def-A1}) the decomposition
for the matrix elements $v_{1\rho}$ and $a_{1\rho}$ in terms of GPDs for
quark species $i$. Here $H_i$ ($\widetilde H_i$) and $E_i$ ($\widetilde
E_i$) are leading twist-two spin non-flip and spin flip GPDs, respectively,
in the parity even (odd) sector. The remaining two sets, containing together
eight independent functions, belong to the twist-three sector. As mentioned
before, the three sets of Compton form factors are given as a convolution of
the coefficient functions with GPDs, generically written as (replace ${\cal
F} \to {\cal F}^3_\pm $ and $F\to F^3_\pm $ to get the result in the
twist-three sector):
\begin{equation}
\label{Def-Con}
{\cal F}
(\xi, \Delta^2)
=
\sum_{i = u, d, \dots}
C^{(\mp)}_i
(\xi, x)
\otimes
F_i
(x, \xi, \Delta^2) .
\end{equation}

In order to deduce the GPDs  from the result in Ref.\
\cite{BelMul00b}, we decompose the (axial-) vector Dirac bilinears $h_\rho$
($\widetilde h_\rho$) in its twist-two and -three components by means of the
Dirac equation
\begin{eqnarray}
\label{Dec-FouVec}
h_{\rho}
\!\!\!&=&\!\!\!
P_\rho \frac{q \cdot h}{q \cdot P}
+ \frac{4 M^2}{(1 - \xi^2)\left( \Delta^2 - \Delta^2_{\rm min} \right)}
\Bigg\{
\Delta_\rho^\perp \xi
\left(
\frac{\Delta^2}{4 M^2} \frac{q \cdot h}{q \cdot P}
-
\frac{q\cdot e}{q \cdot P}
\right)
+ \widetilde\Delta_\rho^\perp
\left(
\frac{\Delta^2}{4M^2} \frac{q \cdot \widetilde h}{q \cdot P}
-
\frac{q \cdot \widetilde e}{q \cdot P}
\right)
\Bigg\},
\nonumber\\
\widetilde h_{\rho}
\!\!\!&=&\!\!\!
P_\rho \frac{q \cdot \widetilde h}{q \cdot P}
+ \frac{4 M^2}{(1 - \xi^2) \left( \Delta^2 - \Delta^2_{\rm min} \right)}
\Bigg\{
\Delta_\rho^\perp \xi
\left(
\left( \frac{\Delta^2}{4 M^2} - 1 \right)
\frac{q \cdot \widetilde h}{q \cdot P}
-  \frac{1}{\xi^2} \frac{q \cdot \widetilde e}{q \cdot P}
\right) \\
&&\quad\qquad\qquad\qquad\qquad\qquad\qquad\qquad\qquad\qquad\qquad\qquad
\qquad \
+ \ \widetilde\Delta_\rho^\perp
\left(
\frac{\Delta^2}{4 M^2} \frac{q \cdot h}{q \cdot P}
-
\frac{q \cdot e}{q \cdot P}
\right)
\Bigg\},
\nonumber
\end{eqnarray}
where twist-four terms, proportional to $q_\rho$, have been neglected. The
GPDs can now be easily read off from Ref.\ \cite{BelMul00b}. All twist-three
GPDs are decomposed in the so-called Wandzura-Wilczeck terms $F^{WW}_\pm$
(also calculated in Ref.\ \cite{KivPol00}) and a function $F^{qGq}_\pm $
that contains new dynamical information arising from antiquark-gluon-quark
correlations:
\begin{eqnarray}
F^3_\pm = F^{WW}_\pm + F^{qGq}_\pm .
\end{eqnarray}
According to the analysis of Ref.\ \cite{BelMul00b}, the WW parts have the
following form
\begin{eqnarray}
\label{Def-Ftw3}
&& F^{WW}_+ =
\frac{1}{\xi} W_+ \otimes \hat d \, F - \frac{1}{\xi} F -
\frac{4 M^2}{(1 - \xi^2)(\Delta^2 - \Delta^2_{\rm min})}
F^\perp_+,
\nonumber\\
&& F^{WW}_- = - \frac{1}{\xi} W_- \otimes \hat d \, F -
\frac{4 M^2}{(1 - \xi^2)(\Delta^2 - \Delta^2_{\rm min})} F^\perp_-,
\end{eqnarray}
where the convolution with the $W$-kernels
\begin{equation}
W_{\pm} \left( \frac{x}{\xi}, \frac{y}{\xi} \right)
= \frac{1}{2 \xi}
\left\{
W \left( \frac{x}{\xi}, \frac{y}{\xi} \right)
\pm
W \left( - \frac{x}{\xi}, - \frac{y}{\xi} \right)
\right\},
\qquad
W(x, y)
= \frac{\theta (1 + x) - \theta (x - y)}{1 + y} ,
\end{equation}
is defined as previously in Eq.\ (\ref{Convolution})
\begin{equation}
W \otimes \hat d \, F
\equiv
\int_{- 1}^{1} d y \,
W \left( \frac{x}{\xi}, \frac{y}{\xi} \right) \hat d (y, \xi) F(y, \xi) ,
\end{equation}
with the differential operator $\hat d (y, \xi) = y \ft{\stackrel{\leftarrow}{
\partial}}{\partial y} - \xi \ft{\stackrel{\rightarrow}{\partial}}{\partial
\xi}$.

Compared to the results for a (pseudo) scalar target (cf. with Eq.\ (24) in
Ref.\ \cite{BelMulKirSch00}), in the WW-sector of Eq.\ (\ref{Def-Ftw3}) there
appear in addition the functions $F_\pm^\perp$, which arise from the
decomposition of $h^\perp_\rho = h_\rho - P_\rho \, q \cdot h / q  \cdot P$ and
$\widetilde h^\perp_\rho = \widetilde h_\rho - P_\rho \, q \cdot \widetilde h /
q \cdot P$, respectively. Their calculation is straightforward and they read
\begin{eqnarray}
H^\perp_\pm \!\!\!&=&\!\!\! \mp \frac{\Delta^2}{4 M^2} \left\{ \xi W_\pm
\otimes \left( H + E \right) - W_\mp \otimes \widetilde H \right\} ,
\nonumber\\
E^\perp_\pm \!\!\!&=&\!\!\! \pm \left\{ \xi W_\pm \otimes \left( H
+ E \right) - W_\mp \otimes \widetilde H \right\},
\nonumber\\
\widetilde
H^\perp_\pm \!\!\!&=&\!\!\! \pm \left\{ \xi \left( 1 - \frac{\Delta^2}{4 M^2}
\right) W_\pm \otimes \widetilde H + \frac{\Delta^2}{4 M^2} W_\mp \otimes
\left( H + E \right) \right\} ,
\nonumber\\
\widetilde E^\perp_\pm
\!\!\!&=&\!\!\! \pm \frac{1}{\xi} \left\{ W_\pm \otimes \widetilde H -
\xi W_\mp \otimes \left( H + E \right) \right\} .
\end{eqnarray}
The antiquark-gluon-quark contributions,
\begin{eqnarray*}
\label{Def-FqGq}
F^{qGq}_\pm
\!\!\!&=&\!\!\!
- \int_{-1}^1 \frac{dy}{\xi}
\int_{-1}^1 du \frac{1-u}{2}
\left\{
W \left( -\frac{x}{\xi}, - \frac{y}{\xi} \right)
\frac{\stackrel{\leftarrow}{\partial^2}}{\partial y^2}
S_F^+(y, u, - \xi)
\pm
W \left( \frac{x}{\xi}, \frac{y}{\xi} \right)
\frac{\stackrel{\leftarrow}{\partial^2}}{\partial y^2}
S_F^-(y,-u,-\xi)
\right\} ,
\end{eqnarray*}
can be read off from the parametrization of the corresponding operators
and result in eight independent functions
\begin{eqnarray*}
S_\rho^{\pm}
= \Delta_\rho^\perp  \frac{q\cdot h}{q\cdot P}
S_H^{\pm}
+
\Delta_\rho^\perp \frac{q\cdot e}{q\cdot P}
S_E^{\pm}
\pm
\widetilde{\Delta}_\rho^\perp  \frac{q\cdot \widetilde{h}}{q\cdot P}
S_{\widetilde H}^{\pm}
\pm
\widetilde{\Delta}_\rho^\perp  \frac{q\cdot \widetilde{e}}{q\cdot P}
S_{\widetilde E}^{\pm}
\end{eqnarray*}
(see Ref.\ \cite{BelMul00b} for details).

It turn out that only the difference $F_+^3 -F_-^3$ will enter in the
DVCS amplitude, thus, only four new GPDs remain at twist-three level.
In the WW approximation, i.e.\ neglecting the quark-gluon-quark correlation,
all the twist-three Compton form factors are entirely determined by the
four twist-two GPDs $H$, $E$, $\widetilde H$ and $\widetilde E$.

{\bf 3.} In this section we define the twist-three observables, which are
accessible by measuring the four-fold cross section for the process $e (k)
h (P_1) \to e (k^\prime) h (P_2) \gamma (q_2)$:
\begin{eqnarray}
\label{WQ}
\frac{d\sigma}{d\Bx dy d|\Delta^2| d\phi}
=
\frac{\alpha^3  \Bx y } { 8 \, \pi \,  {\cal Q}^2}
\left( 1 + \frac{4 M^2 \Bx^2}{{\cal Q}^2} \right)^{-1/2}
\left| \frac{\cal T}{e^3} \right|^2.
\end{eqnarray}
This cross section depends on the Bjorken variable $\Bx = - q_1^2/(2
P_1\cdot q_1)$ (with $q_1 = k - k'$ and $q_1^2 = - {\cal Q}^2$), the
momentum transfer square $\Delta^2 = (P_2 - P_1)^2$, the lepton energy
fraction $y = P_1\cdot q_1/P_1\cdot k$, and the azimuthal angle $\phi$
between lepton and hadron scattering planes. In the twist-three
approximation $\xi$ can be replaced by $\Bx$ via $\xi = \Bx/(2-\Bx)$.
Furthermore, we choose a frame rotated w.r.t.\ the laboratory one in such a
way that the virtual photon has no transverse components\footnote{It is a
reference system related to the centre-of-mass system of
\cite{GouDiePirRal97} by a boost of the hadron in the $z$-direction.}. We
fix our kinematics by choosing a negative $z$ component for the virtual
photon momentum and a positive $x$ component of the incoming electron: $k =
(E, E \sin\theta_e, 0 ,E \cos\theta_e )$, $q_1 = (q_1^0, 0,-|q_1^3|)$, $P_1 =
(M, 0, 0, 0)$ and $P_2 = (E_2, |\mbox{\boldmath$P$}_2| \cos\phi \sin\theta_H,
|\mbox{\boldmath$P$}_2| \sin\phi \sin\theta_H, |\mbox{\boldmath$P$}_2|
\cos\theta_H)$, where $\phi$ is the azimuthal angle between the lepton and
hadron scattering planes.

The amplitude ${\cal T}$ is the sum of the virtual Compton scattering (VCS)
${\cal T}_{VCS}$ and the BH amplitude ${\cal T}_{BH}$. The latter one is
real and is parametrized in terms of electromagnetic form factors, which we
assume to be known from other measurements. The azimuthal angle dependence
of each of the three terms in ${\cal T}^2= |{\cal T}_{BH}|^2 + |{\cal
T}_{VCS}|^2 + {\cal I}$, where ${\cal I} \equiv {\cal T}_{VCS} {\cal
T}_{BH}^\ast + {\cal T}_{VCS}^\ast {\cal T}_{BH}$, is given in our frame by
a finite Fourier sum. In the case of an unplolarized or longitudinally
polarized lepton beam, the interference term ${\cal I}$ and the squared DVCS
amplitude $|{\cal T}_{VCS}|^2$ may be written as
\begin{eqnarray}
&&{\cal I}
=
\frac{1}{\Bx y^3 {\cal P}_1 {\cal P}_2 (-\Delta^2)}
\left\{
\frac{\Delta^2}{{\cal Q}^2}
c_0^{\cal I} + \sum_{m = 1}^2
K^m
\left[ c_m^{\cal I} \cos(m\phi) + \lambda s_m^{\cal I} \sin (m\phi)
\right]
\right\} ,
\nonumber\\
\label{AmplitudesSquared}
&& |{\cal T}_{\rm DVCS}|^2
=
\frac{1}{y^2 {\cal Q}^2}
\left\{
c^{\rm DVCS}_0
+ K \left[ c^{\rm DVCS}_1 \, \cos (\phi)
+ \lambda \, s^{\rm DVCS}_1\, \sin(\phi)
\right]
\right\} .
\end{eqnarray}
The coefficients $c^{\cal I}_1, s^{\cal I}_1$ as well as $c^{\rm DVCS}_0$
and $s^{\rm DVCS}_0$ arise at the twist-two level and their dependence on
GPDs has been calculated in Refs.\ \cite{GouDiePirRal97,BelMueNieSch00}.
$c^{\cal I}_0$, $c^{\cal I}_2$, $s^{\cal I}_2$,
$c^{\rm DVCS}_1$, and $s^{\rm DVCS}_1$ provide an additional angular
dependence and are given in terms of twist-two and twist-three GPDs, while
the terms\footnote{They are induced at twist-two level by the gluon
transversity, which is perturbatively suppressed by $\alpha_s$ and is
contaminated by twist-four contributions.} that are discarded here are
either proportional to $\cos{(3\phi)}$ [$\cos{(2\phi)}$] or $\sin{(3\phi)}$
[$\sin{(2\phi)}$] for the interference [squared DVCS] term. Note that all
$c$'s and $s$'s are $\phi$ independent. There is an
important difference between the interference term and the squared DVCS
amplitude. The former has an additional $\phi$ dependence due to
the lepton propagators\footnote{For convenience we scale them with respect
to ${\cal Q}^2$.}: \begin{equation}
\label{ExaBHpro}
{\cal Q}^2 {\cal P}_1 \equiv (k - q_2)^2 = {\cal Q}^2 + 2k\cdot \Delta,
\qquad
{\cal Q}^2 {\cal P}_2 \equiv (k - \Delta)^2 = - 2 k \cdot \Delta + \Delta^2,
\end{equation}
with
$$
k\cdot \Delta
= - \frac{{\cal Q}^2}{2y}
\Bigg\{
1 + 2 K \cos{\phi} - \frac{\Delta^2}{{\cal Q}^2}
\left( 1 - \Bx (2-y)  \right)
-  \frac{2 M^2}{{\cal Q}^2} (2-y)\Bx^2
\Bigg\}
\left\{ 1 + {\cal O} \left( M^2/{\cal Q}^2,  \Delta^2/{\cal Q}^2 \right)
\right\} .
$$
The $1/{\cal Q}$-power suppressed kinematical factor, appearing also in the
Fourier series (\ref{AmplitudesSquared}),
\begin{equation}
K = \left\{-\frac{\Delta^2}{{\cal Q}^2} (1 - \Bx)
\left( 1 - y \right)
\left( 1 - \frac{\Delta^2_{\rm min}}{\Delta^2} \right) \right\}^{1/2}
\left\{ 1 + {\cal O} \left( M^2/{\cal Q}^2,  \Delta^2/{\cal Q}^2 \right)
\right\}
\end{equation}
vanishes at the kinematical boundary $\Delta^2 = \Delta_{\rm min}^2$,
determined by the minimal value
\begin{eqnarray*}
-\Delta_{\rm min}^2
= \frac{M^2 \Bx^2}{1 - \Bx + \Bx M^2/{\cal Q}^2}
\left\{ 1 + {\cal O} \left( M^2/{\cal Q}^2 \right) \right\} .
\end{eqnarray*}

Since $\cal I$ and $|{\cal T}_{BH}|^2+|{\cal T}_{VCS}|^2$ are charge odd and
even, respectively, those experimental facilities that possess beams of both
charges can separately extract the BH-DVCS interference term and the sum of the
squared BH and DVCS terms. Combining the charge asymmetry with different
nucleon/lepton polarizations and azimuthal asymmetries
\cite{Ji96,GouDiePirRal97, BelMueNieSch00} one can extract, in principle,
the real and imaginary part of the four twist-two amplitudes $\cal H$, $\cal
E$, $\widetilde {\cal H}$, and $\widetilde {\cal E}$ from the interference
term \cite{BelMueNieSch00}. Since ${\cal I}$ is linear in amplitudes
involving GPDs, one gets access to the twist-two GPDs, which, however,
are convoluted with the real or imaginary part of the coefficient functions.
The precise definition of the Fourier coefficients is crucial for the
interpretation of experimental results, since the $u$-channel lepton
propagator gives a contribution that behaves like ${\cal P}_1 = \left[ 1-y -
K \cos\phi +{\cal O} (1/{\cal Q}^2) \right]/y$. For instance, in the case of
the single lepton spin asymmetry it may be possible that in some kinematical
regions, i.e.\ large $y$ and not too small $\Delta^2/ {\cal Q}^2$, this
propagator effectively induces a $\sin(2\phi)$ term that is canceled by a
$s_2^{\cal I} \sin{(2\phi)}$ term resulting in a (fake) $\sin\phi$
dependence of the cross section.

The procedure, outlined above, can also be used to access the twist-three
contributions. For a successful separation of definite twist components, one
should compensate for the strong azimuthal dependence of the lepton propagators
in the BH amplitude. The $\cos (m \phi),\ \sin (m \phi)$ components are e.g.\
unraveled by weighting the cross section with ${\cal P}_1 {\cal P}_2 \left\{
\cos (m \phi) ,\ \sin (m \phi) \right\}$. If this is not done, the magnitude of
the twist-three effects can be judged by the distortion of the leading twist
angular dependence $\left\{ \cos (\phi), \sin (\phi) \right\} {\cal P}_1^{-1}
{\cal P}_2^{-1}$.

In the following we only present the Fourier coefficients to
twist-three accuracy for an unpolarized nucleon target. A complete
analysis will be given elsewhere \cite{BelKirMulSch01b}. A
straightforward computation provides the following analytical
results for the squared DVCS amplitude
\begin{eqnarray}
\label{Res-Mom-DVCS}
c^{\rm DVCS}_{0,{\rm unp}}
\!\!\!&=&\!\!\!
2 ( 2 - 2 y + y^2 )
{\cal C}^{\rm DVCS}_{\rm unp}
\left(
{\cal F},{\cal F}^\ast\right) ,
\\
c^{\rm DVCS}_{1,{\rm unp}}
\!\!\! &=&\!\!\!
8 \frac{2 - y}{2 - \Bx}
{\rm Re} \, {\cal C}^{\rm DVCS}_{\rm unp}
\left(
{\cal F}^{\rm eff},{\cal F}^\ast\right) ,
\nonumber\\
s^{\rm DVCS}_{1,{\rm unp}}
\!\!\!&=&\!\!\!
8 \frac{y}{2 - \Bx}
{\rm Im} \, {\cal C}^{\rm DVCS}_{\rm unp}
\left({\cal F}^{\rm eff}, {\cal F}^\ast\right) .
\nonumber
\end{eqnarray}
Both twist-three coefficients $c^{\rm DVCS}_{1,{\rm unp}}$ and $s^{\rm
DVCS}_{1,{\rm unp}}$ are given as interference of twist-two GPDs with an
`effective' twist-three GPD
\begin{eqnarray}
{\cal F}^{\rm eff} = - 2\xi
\left(
\frac{1}{1 + \xi} {\cal F} + {\cal F}^3_+ - {\cal F}^3_-
\right),
\end{eqnarray}
where ${\cal F}^3_\pm$ are defined in Eqs.\ (\ref{Def-Ftw3}-\ref{Def-FqGq}).
Surprisingly they have the same functional dependence as the leading twist-two
function \cite{BelMueNieSch00}:
\begin{eqnarray}
\label{Res-Mom-Int}
{\cal C}^{\rm DVCS}_{\rm{unp}}
\left(
{\cal F},{\cal F}^\ast
\right)
&\!\!\!=&\!\!\!
\frac{1}{(2 - \Bx)^2}
\Bigg\{
4 (1 - \Bx)
\left(
{\cal H} {\cal H}^\ast
+
\widetilde{\cal H} \widetilde {\cal H}^\ast
\right)- \Bx^2
\bigg(
{\cal H} {\cal E}^\ast
+ {\cal E} {\cal H}^\ast
+ \widetilde{{\cal H}} \widetilde{{\cal E}}^\ast
+ \widetilde{{\cal E}} \widetilde{{\cal H}}^\ast
\bigg)
\nonumber\\
&&\qquad\qquad\;
-
\left( \Bx^2 + (2 - \Bx)^2 \frac{\Delta^2}{4M^2} \right)
{\cal E} {\cal E}^\ast
- \Bx^2 \frac{\Delta^2}{4M^2}
\widetilde{{\cal E}} \widetilde{{\cal E}}^\ast
\Bigg\} .
\end{eqnarray}
For the interference term we found the same property for
the $\cos (2\phi)/\sin (2\phi)$ coefficients:
\begin{eqnarray}
\label{Res-IntTer}
c^{\cal I}_{0,\rm{unp}}
\!\!\!&=&\!\!\!
- 8 (2 - y)
\left\{
(2 - \Bx) (1 - y) - (1 - \Bx)(2 - y)^2
\left( 1 - \frac{\Delta^2_{\rm min}}{\Delta^2} \right)
\right\}
{\rm Re}\,
{\cal C}^{\cal I}_{\rm unp} \left({\cal F}\right)
\nonumber\\
&+&\!\!\! 8 (2 - y) (1 - y) \Bx  (F_1 + F_2)
{\rm Re}
\left\{
\frac{\Bx}{2 - \Bx} ({\cal H} + {\cal E})
+ \widetilde {\cal H}
\right\} ,
\nonumber\\
c^{\cal I}_{1, \rm unp}
\!\!\!&=&\!\!\!
- 8 (2 - 2y + y^2)
{\rm Re} \,
{\cal C}^{\cal I}_{\rm unp}
\left({\cal F} \right),
\qquad
s^{\cal I}_{1, \rm unp}
=
8 y (2-y)
{\rm Im} \,
{\cal C}^{\cal I}_{\rm unp}
\left( {\cal F} \right) ,
\nonumber\\
c^{\cal I}_{2, \rm unp}
\!\!\!&=&\!\!\!
- 16 \frac{2 - y}{2 - \Bx}
{\rm Re}\,
{\cal C}^{\cal I}_{\rm unp}
\left(
{\cal F}^{\rm eff}\right) ,
\qquad\quad
s^{\cal I}_{2, \rm unp}
=
16\frac{y}{2 - \Bx}
{\rm Im} \,
{\cal C}^{\cal I}_{\rm unp}
\left(
{\cal F}^{\rm eff}\right) .
\end{eqnarray}
Here the twist-two function \cite{BelMueNieSch00}
\begin{eqnarray}
{\cal C}^{\cal I}_{\rm unp}
\left( {\cal F} \right)
=
F_1 {\cal H} + \frac{\Bx}{2 - \Bx}
(F_1 + F_2) \widetilde {\cal H}
-
\frac{\Delta^2}{4M^2} F_2 {\cal E}
\end{eqnarray}
also depends on the Dirac and Pauli form factors $F_1$ and $F_2$,
respectively.

Note that twist-three GPDs, having generical discontinuities at $|x| = \xi$,
enter in the Compton form factors in a singularity free combination just as
for a (pseudo) scalar target. The convolutions\footnote{This can be
done by  means of $\int_{-1}^1\frac{dx}{|\eta|}
\frac{1}{\xi-x-i 0} W\left(\frac{x}{\eta},\frac{y}{\eta}\right)=
\frac{{\rm sign}(\eta)}{\eta+y}
\ln\left(\frac{\eta+\xi}{\xi-y-i 0} \right)$ and proper symmetrization.}
of the coefficient functions $C^\mp_i$ with the $W$-kernels make this property
transparent:
\begin{eqnarray}
\label{Res-tw3eff}
{\cal F}^{\rm eff}(\xi) \!\!\!&=&\!\!\!
\frac{2}{1 + \xi}{\cal F} + 2 \xi
\sum_{i=u,d,\dots}
\Bigg\{
\frac{\partial}{\partial \xi}
C^{3(\mp)}_i (\xi, x) \otimes F_i (x, \xi)
+ \frac{4 M^2}{(1 - \xi^2)(\Delta^2 - \Delta^2_{\rm min})}
{\cal F}^\perp_i(\xi)
\\
&&\!\!\! - \int_{-1}^{1} du \,
\frac{1 + u}{\xi + x}
\ln \left( \frac{2 \xi}{\xi - x - i 0} \right)
\frac{\stackrel{\leftarrow}{\partial^2}}{\partial x^2}\otimes
\left( S^+_{{F_i}} (-x, -u, -\xi) - S^-_{{F_i}} (x, u, -\xi) \right)
\Bigg\},
\nonumber
\end{eqnarray}
where we have used a new convention for the coefficient function
\begin{equation}
C_i^{3(\mp)} (\xi, x)
= \frac{Q^2_i}{\xi + x} \ln \frac{2 \xi}{\xi - x - i 0} \mp \{ x \to -x \} ,
\end{equation}
with the $-$ ($+$) sign standing for ${\cal F}= \{{\cal H} (\widetilde {\cal H}),
\ {\cal E} (\widetilde {\cal E})\}$ and
\begin{eqnarray}
{\cal H}^{\perp}_i
\!\!\!&=&\!\!\!\
- \frac{\Delta^2}{4 M^2}
\left\{
\xi C_i^{3(-)} \otimes \left( H_i + E_i \right)
-
C_i^{3(+)} \otimes \widetilde H_i
\right\},
\\
{\cal E}_i^{\perp}
\!\!\!&=&\!\!\!\
\xi C_i^{3(-)} \otimes \left( H_i + E_i \right)
-
C_i^{3(+)} \otimes \widetilde H_i ,
\nonumber\\
\widetilde {\cal H}_i^{\perp }
\!\!\!&=&\!\!\!
\xi \left(1 - \frac{\Delta^2}{4 M^2} \right)
C_i^{3(+)} \otimes \widetilde H_i
+
\frac{\Delta^2}{4 M^2} C_i^{3(-)} \otimes \left( H_i + E_i \right) ,
\nonumber\\
\widetilde {\cal E}_i^{\perp }
\!\!\!&=&\!\!\!
\frac{1}{\xi} \left\{
C_i^{3(+)} \otimes \widetilde H_i
-
\xi C_i^{3(-)} \otimes \left( H_i + E_i \right)
\right\} .
\nonumber
\end{eqnarray}
We should note that the kinematical factor in front of ${\cal
F}^\perp$ drops out in the final results (\ref{Res-Mom-DVCS},\ref{Res-IntTer})
and, therefore, the $1-\Delta^2_{\rm min}/\Delta^2$ dependence in
the Fourier series (\ref{AmplitudesSquared}) will not be altered.
More precisely, the  ${\cal F}^\perp$ dependence in
the `effective' twist-three amplitude (\ref{Res-tw3eff}) generates the
following terms:
\begin{eqnarray}
{\cal C}^{\cal I}_{\rm unp}
\left(
{\cal F}^{\rm eff}\right)
\!\!\!&=&\!\!\!
{\cal C}^{\cal I}_{\rm unp}
\left(
{\cal F}^{\rm eff}, {\cal F}^{\perp} = 0
\right)
+
\frac{2 \Bx}{2 - \Bx} (F_1 + F_2)
\sum_i C_i^{3(+)}\otimes \widetilde H_i,
\\
{\cal C}^{\rm DVCS}_{\rm unp}
\left(
{\cal F}^{\rm eff} , {\cal F}^\ast
\right)
\!\!\!&=&\!\!\!
{\cal C}^{\rm DVCS}_{\rm unp}
\left(
{\cal F}^{\rm eff} , {\cal F}^\ast ; {\cal F}^{\perp} = 0
\right)
\nonumber\\
&+&\!\!\! 2 \Bx
\left(
\sum_i C_i^{3(-)}\otimes \left( E_i + H_i \right)
\right)
\left(
(2 - \Bx) \widetilde{\cal H}^\ast
-
\Bx ( {\cal E} + {\cal H} )^{\ast}
\right)
\nonumber\\
&+&\!\!\! 2 \Bx
\left(
\sum_i C_i^{3(+)}\otimes \widetilde{H}_i
\right)
\left(
(2-\Bx) ({\cal H}+{\cal E})^\ast
-
\Bx (\widetilde{\cal H}+\widetilde{\cal E})^\ast
\right) .
\nonumber
\end{eqnarray}

{\bf 4.} In this paper we have defined twist-two and twist-three observables
in such a way that the twist-three contributions, suppressed by ${\cal O}
\left( \Delta_\perp/{\cal Q} \right)$, do not alter the leading twist
angular dependence. Thus, the corrections to the twist-two Fourier
components are expected at twist-four level ${\cal O} \left( M^2/{\cal Q}^2,
\Delta^2/{\cal Q}^2 \right)$. Combining charge and spin asymmetries
together
with a careful extraction of the different Fourier coefficients allows a
separate measurement of all coefficients of the interference and the squared
DVCS term. In the latter case, the subtraction of the squared BH
amplitude from the data is necessary. At the same time it drops out in
single spin asymmetries.

For a lepton beam with definite charge, a separate measurement of the
interference term is not possible. In this case the leading twist $\cos (\phi)/
\sin (\phi)$ dependence in the cross section, which stems from the interference
term, gets corrected by the twist-three contribution of the squared DVCS
amplitude. Although such corrections are kinematically suppressed by $\Bx
\ft{1 - y}{y} \Delta^2/{\cal Q}^2$, their numerical value depends on the
magnitude of the form factors $\cal F$.

Let us also mention that for a (pseudo) scalar target the Fourier
coefficients fulfill certain constraints derived from general arguments
\cite{BelMulKirSch00}. They provide a basic test of the applicability of the
operator product expansion at a given $\cal Q$-scale. We expect similar
constraints for a spin-1/2 target, however, their derivation will require
an evaluation of the squared amplitudes for longitudinally and transversely
polarized targets. Of course, one can not test the dominance of leading
contributions in this way or the magnitude of higher twist and $\alpha_s$
suppressed contaminations. This can be judged by comparing the measured
scale dependence of the coefficients $c_i$ and $s_i$ with the ones arising
from the evolution of GPDs and power suppressed contributions in a sufficiently
large ${\cal Q}^2$-interval. Note that this task is complicated for experiments
with fixed center-of-mass energy, since the kinematical variables $\Bx, y$ and
$\cal Q$ do not form an independent basis in this setting. On the
theoretical side the complete perturbative NLO corrections have been
calculated while an estimate of twist-four effects still has to be done.

Extraction of the leading twist-two and -three Fourier coefficients allows
to test models for GPDs and even to measure them in single spin asymmetries
on the diagonal\footnote{This statement is only true at leading order in
perturbation theory. Unfortunately, there are large pertubative corrections
that require a careful interpretation of such experimental results.}
$x = \xi$. Having tested those predictions and processing to higher twist, one
can study the validity of the `Wandzura-Wilczek model' for twist-three
functions suggested in Ref.\ \cite{BelMul00b}. A significant deviation from
the latter would imply an essential contribution from antiquark-gluon-quark
correlators.

To conclude, the full experimental exploration of the deep electroproduction
of real photons with leptons of both charges and polarizations will lead
to a direct confrontation of the data with theoretical predictions and will
result in systematic tests of our understanding of the quark-gluon
content of the nucleon via generalized parton distributions.

\end{document}